\PassOptionsToPackage{unicode}{hyperref}
\PassOptionsToPackage{hyphens}{url}
\PassOptionsToPackage{dvipsnames,svgnames,x11names}{xcolor}
\documentclass[
  letterpaper,
  DIV=11,
  numbers=noendperiod]{scrartcl}

\usepackage{amsmath,amssymb}
\usepackage{iftex}
\ifPDFTeX
  \usepackage[T1]{fontenc}
  \usepackage[utf8]{inputenc}
  \usepackage{textcomp} 
\else 
  \usepackage{unicode-math}
  \defaultfontfeatures{Scale=MatchLowercase}
  \defaultfontfeatures[\rmfamily]{Ligatures=TeX,Scale=1}
\fi
\usepackage{lmodern}
\ifPDFTeX\else  
\fi
\IfFileExists{upquote.sty}{\usepackage{upquote}}{}
\IfFileExists{microtype.sty}{
  \usepackage[]{microtype}
  \UseMicrotypeSet[protrusion]{basicmath} 
}{}
\makeatletter
\@ifundefined{KOMAClassName}{
  \IfFileExists{parskip.sty}{%
    \usepackage{parskip}
  }{
    \setlength{\parindent}{0pt}
    \setlength{\parskip}{6pt plus 2pt minus 1pt}}
}{
  \KOMAoptions{parskip=half}}
\makeatother
\usepackage{xcolor}
\ifx\paragraph\undefined\else
  \let\oldparagraph\paragraph
  \renewcommand{\paragraph}[1]{\oldparagraph{#1}\mbox{}}
\fi
\ifx\subparagraph\undefined\else
  \let\oldsubparagraph\subparagraph
  \renewcommand{\subparagraph}[1]{\oldsubparagraph{#1}\mbox{}}
\fi

\usepackage{color}
\usepackage{fancyvrb}

\DefineVerbatimEnvironment{Highlighting}{Verbatim}{commandchars=\\\{\}}
\usepackage{framed}
\definecolor{shadecolor}{RGB}{241,243,245}
\newenvironment{Shaded}{\begin{snugshade}}{\end{snugshade}}

\newcommand{\BuiltInTok}[1]{\textcolor[rgb]{0.00,0.23,0.31}{#1}}
\newcommand{\CharTok}[1]{\textcolor[rgb]{0.13,0.47,0.30}{#1}}
\newcommand{\CommentTok}[1]{\textcolor[rgb]{0.37,0.37,0.37}{#1}}

\newcommand{\DecValTok}[1]{\textcolor[rgb]{0.68,0.00,0.00}{#1}}

\newcommand{\FloatTok}[1]{\textcolor[rgb]{0.68,0.00,0.00}{#1}}

\newcommand{\ImportTok}[1]{\textcolor[rgb]{0.00,0.46,0.62}{#1}}
\newcommand{\InformationTok}[1]{\textcolor[rgb]{0.37,0.37,0.37}{#1}}

\newcommand{\NormalTok}[1]{\textcolor[rgb]{0.00,0.23,0.31}{#1}}
\newcommand{\OperatorTok}[1]{\textcolor[rgb]{0.37,0.37,0.37}{#1}}

\newcommand{\StringTok}[1]{\textcolor[rgb]{0.13,0.47,0.30}{#1}}

\providecommand{\tightlist}{%
  \setlength{\itemsep}{0pt}\setlength{\parskip}{0pt}}\usepackage{longtable,booktabs,array}
\usepackage{calc} 
\usepackage{etoolbox}
\makeatletter
\patchcmd\longtable{\par}{\if@noskipsec\mbox{}\fi\par}{}{}
\makeatother
\IfFileExists{footnotehyper.sty}{\usepackage{footnotehyper}}{\usepackage{footnote}}
\makesavenoteenv{longtable}
\usepackage{graphicx}
\makeatletter
\def\maxwidth{\ifdim\Gin@nat@width>\linewidth\linewidth\else\Gin@nat@width\fi}
\def\maxheight{\ifdim\Gin@nat@height>\textheight\textheight\else\Gin@nat@height\fi}
\makeatother
\setkeys{Gin}{width=\maxwidth,height=\maxheight,keepaspectratio}
\makeatletter
\def\fps@figure{htbp}
\makeatother
\newlength{\cslhangindent}
\newlength{\csllabelwidth}
\newlength{\cslentryspacingunit} 
\newenvironment{CSLReferences}[2] 
 {
  \setlength{\parindent}{0pt}
  \ifodd #1
  \let\oldpar\par
  \def\par{\hangindent=\cslhangindent\oldpar}
  \fi
  \setlength{\parskip}{#2\cslentryspacingunit}
 }%
 {}
\usepackage{calc}

\KOMAoption{captions}{tableheading}
\makeatletter
\@ifpackageloaded{caption}{}{\usepackage{caption}}
\AtBeginDocument{%
\ifdefined\contentsname
  \renewcommand*\contentsname{Table of contents}
\else
  \newcommand\contentsname{Table of contents}
\fi
\ifdefined\listfigurename
  \renewcommand*\listfigurename{List of Figures}
\else
  \newcommand\listfigurename{List of Figures}
\fi
\ifdefined\listtablename
  \renewcommand*\listtablename{List of Tables}
\else
  \newcommand\listtablename{List of Tables}
\fi
\ifdefined\figurename
  \renewcommand*\figurename{Figure}
\else
  \newcommand\figurename{Figure}
\fi
\ifdefined\tablename
  \renewcommand*\tablename{Table}
\else
  \newcommand\tablename{Table}
\fi
}
\@ifpackageloaded{float}{}{\usepackage{float}}
\floatstyle{ruled}
\@ifundefined{c@chapter}{\newfloat{codelisting}{h}{lop}}{\newfloat{codelisting}{h}{lop}[chapter]}
\floatname{codelisting}{Listing}

\makeatother
\makeatletter
\@ifpackageloaded{caption}{}{\usepackage{caption}}
\@ifpackageloaded{subcaption}{}{\usepackage{subcaption}}
\makeatother
\makeatletter
\@ifpackageloaded{tcolorbox}{}{\usepackage[skins,breakable]{tcolorbox}}
\makeatother
\makeatletter
\@ifundefined{shadecolor}{\definecolor{shadecolor}{rgb}{.97, .97, .97}}
\makeatother
\ifLuaTeX
  \usepackage{selnolig}  
\fi
\IfFileExists{bookmark.sty}{\usepackage{bookmark}}{\usepackage{hyperref}}
\IfFileExists{xurl.sty}{\usepackage{xurl}}{} 
\hypersetup{
  pdftitle={Explaining AI in Finance: Past, Present, Prospects},
  pdfauthor={Barry Quinn},
  colorlinks=true,
  linkcolor={blue},
  filecolor={Maroon},
  citecolor={Blue},
  urlcolor={Blue},
  pdfcreator={LaTeX via pandoc}}

\title{Explaining AI in Finance: Past, Present, Prospects}
\author{Barry Quinn}
\date{}

\begin{document}
\maketitle
\begin{abstract}
This paper explores the journey of AI in finance, with a particular
focus on the crucial role and potential of Explainable AI (XAI). We
trace AI's evolution from early statistical methods to sophisticated
machine learning, highlighting XAI's role in popular financial
applications. The paper underscores the superior interpretability of
methods like Shapley values compared to traditional linear regression in
complex financial scenarios. It emphasizes the necessity of further XAI
research, given forthcoming EU regulations. The paper demonstrates,
through simulations, that XAI enhances trust in AI systems, fostering
more responsible decision-making within finance.
\end{abstract}
\ifdefined\Shaded\renewenvironment{Shaded}{\begin{tcolorbox}[boxrule=0pt, borderline west={3pt}{0pt}{shadecolor}, interior hidden, breakable, enhanced, sharp corners, frame hidden]}{\end{tcolorbox}}\fi

\hypertarget{introduction}{%
\section{Introduction}\label{introduction}}

The rapid digital transformation of the finance industry over the past
few decades has been predominantly driven by the integration of
Artificial Intelligence (AI) and machine learning technologies. These
technologies have heralded a new era in finance, catalysing innovations
in trading, risk management, fraud detection, customer service, and many
other areas, bringing significant changes to business models,
operations, and services (Arner, Barberis, and Buckley 2020). Today,
financial institutions leverage these advanced technologies to generate
insights, automate processes, and improve decision-making.

Despite the revolutionary potential of AI, its application in finance is
not devoid of challenges. For example, a significant issue arises around
the transparency and interpretability of AI decision-making, often
described as the `black box' problem. This term refers to the difficulty
in understanding how complex AI and machine learning models arrive at
their decisions (Dhar 2018). This opacity presents substantial ethical,
legal, and practical challenges, especially in an industry as regulated
and risk-averse as finance (Bhatt, Xiang, and Sharma 2020).

In response to these concerns, Explainable Artificial Intelligence (XAI)
has emerged to make AI's decision-making process more transparent and
understandable to human users (Molnar 2020). This development is
particularly crucial in the financial sector, where understanding the
rationale behind decisions can have enormous implications for trust,
compliance, risk management, and customer satisfaction.

This paper aims to critically analyse the role of XAI in finance,
tracing its historical development, examining its current applications,
and exploring its prospects. By providing a comprehensive review of XAI
in the context of finance, we hope to shed light on its importance and
potential for the industry while addressing ongoing challenges and areas
for future research.

\hypertarget{literature-review}{%
\section{Literature review}\label{literature-review}}

\hypertarget{past-early-application-of-ai-in-finance}{%
\subsection{Past: Early Application of AI in
Finance}\label{past-early-application-of-ai-in-finance}}

The integration of AI in finance has a history that dates back to the
latter half of the 20th century. Initially, financial institutions
deployed rule-based systems for various functions, such as automated
trading and risk analysis. These were the earliest forms of AI applied
to finance and were quite simplistic compared to today's advanced
systems.

These early AI systems were based on sets of pre-programmed rules, often
developed by human experts, and were typically deterministic. Given the
same input, these systems would always provide the same output. They
were transparent and easily interpretable because they followed clearly
defined, pre-set rules (Gomber et al. 2018).

However, the effectiveness of rule-based systems was limited by their
rigid, inflexible design. These systems were not designed to learn from
new data or adapt to changing conditions, making them less useful in the
dynamic world of finance, which is characterised by evolving markets,
changing regulatory landscapes, and unpredictable economic conditions
(Dhar 2018).

The desire for more adaptive, responsive systems led to the developing
of machine learning algorithms. Machine learning represented a
significant advancement in the field of AI. Unlike rule-based systems,
machine learning models could learn from data, identify patterns, and
make predictions or decisions based on those patterns. Machine learning
algorithms found applications in various areas of finance, from
predicting stock prices to identifying fraudulent transactions (Athey
2021).

The transition from rule-based systems to machine learning models marked
a pivotal shift in AI's role in finance. However, this transition was
not without its challenges. Machine learning models, particularly more
complex ones like neural networks, introduced a level of opacity and
complexity that made it difficult for human users to understand how they
made decisions (Rudin 2019). This lack of transparency and
interpretability in machine learning systems became known as the `black
box' problem and formed the backdrop against which the field of
explainable AI emerged.

In summary, the past of AI in finance was marked by the transition from
transparent but inflexible rule-based systems to powerful but opaque
machine learning models. The need to address the transparency issues
introduced by machine learning has led to the development of explainable
AI, the current and future state of which will be discussed in the
following sections.

\hypertarget{present-the-rise-of-machine-learning-and-xai}{%
\subsection{Present: The Rise of Machine Learning and
XAI}\label{present-the-rise-of-machine-learning-and-xai}}

In the present landscape, machine learning models, such as neural
networks, decision trees, and support vector machines, have become
integral parts of financial institutions. These models perform various
tasks, including credit scoring, fraud detection, algorithmic trading,
portfolio optimisation, and customer segmentation (Athey 2021).

Despite their efficiency and sophistication, these models often work as
`black boxes,' where the internal decision-making process is obscured
from the users. This opaqueness can pose considerable challenges. For
example, it hinders human users, such as loan officers or portfolio
managers, from understanding and trusting the model's decisions. In
addition, it poses problems for accountability and compliance on a
regulatory level, especially in jurisdictions where decisions affecting
individuals must be explainable (Rudin 2019).

Enter the Explainable Artificial Intelligence (XAI) field, which seeks
to make AI's decision-making process more transparent and interpretable.
Various techniques fall under the umbrella of XAI, and these can be
broadly classified into model-specific methods and model-agnostic
methods (Molnar 2020).

Model-specific methods, such as coefficient interpretation in linear
regression or feature importance in decision trees, provide insights
into how these specific models operate. However, their application is
limited to the particular model types for which they were developed
(Molnar 2020).

Model-agnostic methods, on the other hand, can be applied to any
machine-learning model. They seek to provide explanations for individual
predictions regardless of the complexity or type of the underlying
model. Examples of these techniques include Local Interpretable
Model-Agnostic Explanations (LIME) and SHapley Additive exPlanations
(SHAP). LIME offers explanations by approximating the prediction of a
complex model with a simpler, locally-fitted model around the prediction
point (Ribeiro, Singh, \& Guestrin, 2016). Meanwhile, SHAP allocates
each feature's contribution to the prediction for individual data points
based on concepts from cooperative game theory (Lundberg and Lee 2017).

Yet, despite these advancements, achieving true explainability in AI
remains a significant challenge. Many of these methods provide post-hoc
explanations, which attempt to interpret the model's behaviour after
training. This process often involves a trade-off between accuracy and
interpretability, with more complex models offering greater accuracy but
less interpretability (Bhatt, Xiang, and Sharma 2020).

Moreover, explainability is not just a technical problem but also a
human-centred one. The effectiveness of an explanation largely depends
on the recipient's perspective and the context in which it is given
(\textbf{Mill?}) (Miller, 2019). For example, a satisfactory explanation
to a data scientist might be incomprehensible to a loan officer or a
customer, indicating that the development of XAI needs to consider the
human factors of understandability and trust.

In conclusion, the present state of XAI in finance is marked by
considerable advancements and ongoing challenges. The shift towards more
transparent and interpretable AI models is underway, with various
methods being developed and applied. However, balancing models'
complexity (and, thereby, performance) and their interpretability
remains a significant hurdle. As we look towards the future, it is
crucial that these challenges are addressed and XAI continues to evolve
to meet the demands of transparency and interpretability in the
financial industry.

\hypertarget{prospects-the-future-of-xai-in-finance}{%
\subsection{Prospects: The Future of XAI in
Finance}\label{prospects-the-future-of-xai-in-finance}}

As we move into the future, the demand for transparency and
interpretability in AI systems within the finance sector is expected to
grow. The future advancements and challenges in the field of XAI reflect
this.

One significant direction for future research and development is
integrating explainability directly into the model-building process
rather than treating it as an afterthought. This approach, often called
intrinsic explainability, involves building naturally interpretable
models, such as explainable boosting machines or interpretable decision
sets (Lakkaraju, Bach, and Leskovec 2016; Lou, Caruana, and Gehrke
2012). Developing such models can help mitigate the trade-off between
accuracy and interpretability that characterises post-hoc explanation
methods.

Furthermore, as the field of XAI evolves, it will be essential to focus
on the users' perspective. What constitutes a `good' explanation can
vary based on the recipient and the context. Therefore, future XAI
methods should tailor answers to different users' needs and
capabilities. They should also address how to communicate these
explanations best to ensure they are understandable and valuable
(Miller, 2019).

Moreover, as AI and XAI become more commonplace in finance, regulations
will likely evolve to address their new challenges. For example, the
European Union's General Data Protection Regulation (GDPR) has already
introduced a `right to explanation', where individuals can ask for
explanations of decisions made by automated systems that affect them
(Goodman and Flaxman 2017). In the future, we might see more regulations
requiring financial institutions to provide clear, understandable
explanations for AI-based decisions.

However, implementing such regulations comes with its challenges. For
example, regulators will need to define what constitutes an
`explanation' and a `decision' in the context of AI. They will also need
to set standards for how detailed and understandable these explanations
must be (Edwards and Veale 2017).

In summary, the future of XAI in finance is ripe with opportunities for
making AI decision-making more transparent and accountable. However, it
also presents challenges that must be addressed through continued
research, development, and thoughtful regulation.

\hypertarget{confronting-econometrics-with-xai}{%
\section{Confronting Econometrics with
XAI}\label{confronting-econometrics-with-xai}}

Explainable AI (XAI) could undoubtedly play a significant role in
improving the field of econometrics, which is the application of
statistical methods to economic data to give empirical content to
economic relations.

Traditionally, econometric models have been designed to be inherently
interpretable, as they often depend on linear relationships and other
simplistic assumptions to ensure that the model's parameters can be
easily interpreted. However, these assumptions can be limiting, as they
might not fully capture the complexities of real-world economic
phenomena.

With the advent of machine learning, econometricians have created models
that can learn complex patterns from data, leading to more accurate
predictions. But the drawback is that these models are often `black
boxes,' making it difficult to understand how they make decisions.

This is where XAI comes in. By using techniques developed under the
umbrella of XAI, econometricians could make their machine-learning
models more transparent, allowing them to understand how each input
variable contributes to the model's predictions. This increased
transparency could make these models more acceptable to economists and
policymakers, who need to understand the decision-making process to make
informed decisions.

For instance, SHapley Additive exPlanations (SHAP) and Local
Interpretable Model-agnostic Explanations (LIME) could be applied to
make machine learning models more interpretable. These techniques
explain individual predictions, helping to understand the contributions
of different features to the model's output.

In addition to improving model transparency, XAI could help address some
statistical challenges in econometrics. For example, XAI could help
econometricians understand the variable importance in their models,
which could help address issues related to multicollinearity, where
independent variables in a regression model are highly correlated.

In conclusion, XAI holds considerable potential in improving
econometrics, mainly as the field increasingly incorporates machine
learning models. Furthermore, by making these models more transparent,
XAI can help econometricians and policymakers better understand and
trust the predictions derived from these models, leading to better
decision-making.

\hypertarget{shapley-values-versus-ols-regression-coefficients}{%
\subsection{Shapley values versus OLS regression
coefficients}\label{shapley-values-versus-ols-regression-coefficients}}

In a linear regression model, let's explore the analogy between Shapley
values and the coefficients.

In a linear regression model, each predictor variable is assigned a
coefficient representing its partial effect on the outcome variable,
controlling for all other predictors. The coefficient can be interpreted
as the expected change in the outcome variable for a one-unit change in
the predictor, holding all other predictors constant.

Analogously, the Shapley value for a player in a cooperative game
represents the average contribution of the player to the worth of all
possible coalitions that the player can be a part of. The Shapley value
considers all possible ways the coalition can be formed and averages
over them.

In both cases, the goal is to fairly distribute some total quantity (the
total worth of the grand coalition in a cooperative game or the total
variance of the outcome variable in a linear regression model) among
different contributors (the players in a cooperative game, or the
predictor variables in a linear regression model).

However, there are significant differences as well. While linear
regression assumes a specific linear and additive form for the
relationship between predictors and the outcome, Shapley values make no
such assumption. Shapley values can handle any type of game, including
non-cooperative games and games with complex interactions between
players.

Also, the computation of Shapley values considers all possible orders in
which players can join the coalition, reflecting the idea that a
player's contribution may depend on which other players are already in
the coalition. In contrast, linear regression coefficients are typically
computed using a method (like ordinary least squares) that does not
consider different orders of entering the predictors into the model.

In the context of explainable AI, the Shapley value concept has been
applied to machine learning models to compute the contribution of each
feature to the prediction for a particular instance. This can provide
more nuanced and reliable interpretations than simply looking at the
coefficients of a linear model, especially for complex models that
capture non-linear and interactive effects.

\hypertarget{coalition-game-and-shapley-values}{%
\subsubsection{Coalition game and Shapley
values}\label{coalition-game-and-shapley-values}}

Let's consider a simple cooperative game involving three players: A, B,
and C, to fix ideas. A characteristic function v gives the worth of each
coalition of players:

\begin{itemize}
\tightlist
\item
  v(\{\}) = 0 (worth of the empty coalition)
\item
  v(\{A\}) = 100
\item
  v(\{B\}) = 200
\item
  v(\{C\}) = 300
\item
  v(\{A, B\}) = 400
\item
  v(\{A, C\}) = 500
\item
  v(\{B, C\}) = 600
\item
  v(\{A, B, C\}) = 800
\end{itemize}

We want to distribute the total worth of the grand coalition (v(\{A, B,
C\}) = 800) among the players in a way that reflects their contribution
to the coalition.

The Shapley value is one way to do this. For each player, it computes
the average marginal contribution of the player to all possible
coalitions. This is done by considering all permutations of the players
and, for each permutation, adding up the marginal contributions of the
player when they join the coalition.

\hypertarget{confronting-regression-coefficients-using-xai}{%
\subsubsection{Confronting regression coefficients using
XAI}\label{confronting-regression-coefficients-using-xai}}

In a linear regression model, each predictor variable is assigned a
coefficient representing its partial effect on the outcome variable,
controlling for all other predictors. The coefficient can be interpreted
as the expected change in the outcome variable for a one-unit change in
the predictor, holding all other predictors constant.

Analogously, the Shapley value for a player in a cooperative game
represents the average contribution of the player to the worth of all
possible coalitions that the player can be a part of. The Shapley value
considers all possible ways the coalition can be formed and averages
over them.

In both cases, the goal is to fairly distribute some total quantity (the
total worth of the grand coalition in a cooperative game or the total
variance of the outcome variable in a linear regression model) among
different contributors (the players in a cooperative game, or the
predictor variables in a linear regression model).

However, there are significant differences as well. While linear
regression assumes a specific linear and additive form for the
relationship between predictors and the outcome, Shapley values make no
such assumption. Shapley values can handle any type of game, including
non-cooperative games and games with complex interactions between
players.

Also, the computation of Shapley values considers all possible orders in
which players can join the coalition, reflecting the idea that a
player's contribution may depend on which other players are already in
the coalition. In contrast, linear regression coefficients are typically
computed using a method (like ordinary least squares) that does not
consider different orders of entering the predictors into the model.

In the context of explainable AI, the Shapley value concept has been
applied to machine learning models to compute the contribution of each
feature to the prediction for a particular instance. This can provide
more nuanced and reliable interpretations than simply looking at the
coefficients of a linear model, especially for complex models that
capture non-linear and interactive effects.

\hypertarget{a-finance-example}{%
\subsubsection{A finance example}\label{a-finance-example}}

To Steelman, the case here is a simulated example. We'll use the
\texttt{sklearn} and \texttt{shap} libraries in Python to fit a linear
regression model and a more complex model (Random Forest) to the
simulated data. Then we'll use the shap library to compute SHAP values
for the Random Forest model.

Let's consider a simple three-factor model (size, value, and momentum
factors) to predict asset returns.

First, let's install the necessary packages (if not already installed):

Now let's import the necessary packages and generate some simulated
data: g

\begin{Shaded}
\begin{Highlighting}[]
\InformationTok{\textasciigrave{}\textasciigrave{}\textasciigrave{}\{python\}}
\ImportTok{import}\NormalTok{ numpy }\ImportTok{as}\NormalTok{ np}
\ImportTok{import}\NormalTok{ pandas }\ImportTok{as}\NormalTok{ pd}
\ImportTok{from}\NormalTok{ sklearn.linear\_model }\ImportTok{import}\NormalTok{ LinearRegression}
\ImportTok{from}\NormalTok{ sklearn.ensemble }\ImportTok{import}\NormalTok{ RandomForestRegressor}
\ImportTok{import}\NormalTok{ warnings}
\NormalTok{warnings.filterwarnings(}\StringTok{"ignore"}\NormalTok{)}
\ImportTok{import}\NormalTok{ shap}

\CommentTok{\# Set a seed for reproducibility}
\NormalTok{np.random.seed(}\DecValTok{0}\NormalTok{)}

\CommentTok{\# Generate simulated factor values}
\NormalTok{n }\OperatorTok{=} \DecValTok{1000}  \CommentTok{\# number of assets}
\NormalTok{size }\OperatorTok{=}\NormalTok{ np.random.normal(}\DecValTok{0}\NormalTok{, }\DecValTok{1}\NormalTok{, n)}
\NormalTok{value }\OperatorTok{=}\NormalTok{ np.random.normal(}\DecValTok{0}\NormalTok{, }\DecValTok{1}\NormalTok{, n)}
\NormalTok{momentum }\OperatorTok{=}\NormalTok{ np.random.normal(}\DecValTok{0}\NormalTok{, }\DecValTok{1}\NormalTok{, n)}

\CommentTok{\# Generate simulated asset returns}
\CommentTok{\# The true model has coefficients [0.1, 0.2, 0.3] for the factors}
\CommentTok{\# We also add some noise}
\NormalTok{returns }\OperatorTok{=} \FloatTok{0.1}\OperatorTok{*}\NormalTok{size }\OperatorTok{+} \FloatTok{0.2}\OperatorTok{*}\NormalTok{value }\OperatorTok{+} \FloatTok{0.3}\OperatorTok{*}\NormalTok{momentum }\OperatorTok{+}\NormalTok{ np.random.normal(}\DecValTok{0}\NormalTok{, }\FloatTok{0.1}\NormalTok{, n)}

\CommentTok{\# Put data in a DataFrame}
\NormalTok{data }\OperatorTok{=}\NormalTok{ pd.DataFrame(\{}
    \StringTok{\textquotesingle{}Size\textquotesingle{}}\NormalTok{: size,}
    \StringTok{\textquotesingle{}Value\textquotesingle{}}\NormalTok{: value,}
    \StringTok{\textquotesingle{}Momentum\textquotesingle{}}\NormalTok{: momentum,}
    \StringTok{\textquotesingle{}Return\textquotesingle{}}\NormalTok{: returns}
\NormalTok{\})}
\InformationTok{\textasciigrave{}\textasciigrave{}\textasciigrave{}}
\end{Highlighting}
\end{Shaded}

Next, let's fit a linear regression model to the data:

\begin{Shaded}
\begin{Highlighting}[]
\InformationTok{\textasciigrave{}\textasciigrave{}\textasciigrave{}\{python\}}
\NormalTok{X }\OperatorTok{=}\NormalTok{ data[[}\StringTok{\textquotesingle{}Size\textquotesingle{}}\NormalTok{, }\StringTok{\textquotesingle{}Value\textquotesingle{}}\NormalTok{, }\StringTok{\textquotesingle{}Momentum\textquotesingle{}}\NormalTok{]]}
\NormalTok{y }\OperatorTok{=}\NormalTok{ data[}\StringTok{\textquotesingle{}Return\textquotesingle{}}\NormalTok{]}

\NormalTok{lin\_reg }\OperatorTok{=}\NormalTok{ LinearRegression().fit(X, y)}

\CommentTok{\# Print coefficients}
\BuiltInTok{print}\NormalTok{(}\StringTok{\textquotesingle{}Linear Regression Coefficients:\textquotesingle{}}\NormalTok{)}
\BuiltInTok{print}\NormalTok{(}\StringTok{\textquotesingle{}Size:\textquotesingle{}}\NormalTok{, lin\_reg.coef\_[}\DecValTok{0}\NormalTok{])}
\BuiltInTok{print}\NormalTok{(}\StringTok{\textquotesingle{}Value:\textquotesingle{}}\NormalTok{, lin\_reg.coef\_[}\DecValTok{1}\NormalTok{])}
\BuiltInTok{print}\NormalTok{(}\StringTok{\textquotesingle{}Momentum:\textquotesingle{}}\NormalTok{, lin\_reg.coef\_[}\DecValTok{2}\NormalTok{])}
\InformationTok{\textasciigrave{}\textasciigrave{}\textasciigrave{}}
\end{Highlighting}
\end{Shaded}

\begin{verbatim}
Linear Regression Coefficients:
Size: 0.10091701879836647
Value: 0.20185690366007059
Momentum: 0.30119085670658874
\end{verbatim}

Now let's fit a Random Forest model to the data and compute SHAP values:

\begin{Shaded}
\begin{Highlighting}[]
\InformationTok{\textasciigrave{}\textasciigrave{}\textasciigrave{}\{python\}}
\CommentTok{\# Fit model}
\NormalTok{rf }\OperatorTok{=}\NormalTok{ RandomForestRegressor(n\_estimators}\OperatorTok{=}\DecValTok{100}\NormalTok{, random\_state}\OperatorTok{=}\DecValTok{0}\NormalTok{).fit(X, y)}

\CommentTok{\# Explain model predictions using SHAP}
\NormalTok{explainer }\OperatorTok{=}\NormalTok{ shap.Explainer(rf, X)}
\NormalTok{shap\_values }\OperatorTok{=}\NormalTok{ explainer(X)}

\CommentTok{\# Print mean absolute SHAP values for each feature}
\BuiltInTok{print}\NormalTok{(}\StringTok{\textquotesingle{}}\CharTok{\textbackslash{}n}\StringTok{Mean Absolute SHAP Values:\textquotesingle{}}\NormalTok{)}
\BuiltInTok{print}\NormalTok{(}\StringTok{\textquotesingle{}Size:\textquotesingle{}}\NormalTok{, np.mean(np.}\BuiltInTok{abs}\NormalTok{(shap\_values.values[:, }\DecValTok{0}\NormalTok{])))}
\BuiltInTok{print}\NormalTok{(}\StringTok{\textquotesingle{}Value:\textquotesingle{}}\NormalTok{, np.mean(np.}\BuiltInTok{abs}\NormalTok{(shap\_values.values[:, }\DecValTok{1}\NormalTok{])))}
\BuiltInTok{print}\NormalTok{(}\StringTok{\textquotesingle{}Momentum:\textquotesingle{}}\NormalTok{, np.mean(np.}\BuiltInTok{abs}\NormalTok{(shap\_values.values[:, }\DecValTok{2}\NormalTok{])))}
\InformationTok{\textasciigrave{}\textasciigrave{}\textasciigrave{}}
\end{Highlighting}
\end{Shaded}

\begin{verbatim}

 90%|==================  | 897/1000 [00:11<00:01]       
 98%|===================| 979/1000 [00:12<00:00]       

Mean Absolute SHAP Values:
Size: 0.0662061209562134
Value: 0.15558826181291682
Momentum: 0.23316567312446296
\end{verbatim}

In this simulated example, the actual model is linear, so the linear
regression coefficients should be close to the actual values, and the
mean absolute SHAP values for the Random Forest model should also
reflect the relative importance of the factors.

However, in real-world scenarios where the relationship between factors
and asset returns might be non-linear or involve interactions, the
Random Forest model could provide better predictive accuracy, and the
SHAP values could offer a more nuanced understanding of feature
contributions.

Let's modify the above example to include non-linear and interaction
effects. In this modified example, the true polynomial model includes
square terms and an interaction term.

\begin{Shaded}
\begin{Highlighting}[]
\InformationTok{\textasciigrave{}\textasciigrave{}\textasciigrave{}\{python\}}
\CommentTok{\# Generate simulated asset returns with non{-}linear and interaction effects}
\NormalTok{returns }\OperatorTok{=} \FloatTok{0.1}\OperatorTok{*}\NormalTok{size }\OperatorTok{+} \FloatTok{0.2}\OperatorTok{*}\NormalTok{value }\OperatorTok{+} \FloatTok{0.3}\OperatorTok{*}\NormalTok{momentum }\OperatorTok{+} \FloatTok{0.4}\OperatorTok{*}\NormalTok{size}\OperatorTok{**}\DecValTok{2} \OperatorTok{+} \FloatTok{0.5}\OperatorTok{*}\NormalTok{value}\OperatorTok{**}\DecValTok{2} \OperatorTok{+} \FloatTok{0.6}\OperatorTok{*}\NormalTok{momentum}\OperatorTok{**}\DecValTok{2} \OperatorTok{+} \FloatTok{0.7}\OperatorTok{*}\NormalTok{size}\OperatorTok{*}\NormalTok{value }\OperatorTok{+}\NormalTok{ np.random.normal(}\DecValTok{0}\NormalTok{, }\FloatTok{0.1}\NormalTok{, n)}

\CommentTok{\# Put data in a DataFrame}
\NormalTok{data }\OperatorTok{=}\NormalTok{ pd.DataFrame(\{}
    \StringTok{\textquotesingle{}Size\textquotesingle{}}\NormalTok{: size,}
    \StringTok{\textquotesingle{}Value\textquotesingle{}}\NormalTok{: value,}
    \StringTok{\textquotesingle{}Momentum\textquotesingle{}}\NormalTok{: momentum,}
    \StringTok{\textquotesingle{}Return\textquotesingle{}}\NormalTok{: returns}
\NormalTok{\})}

\NormalTok{X }\OperatorTok{=}\NormalTok{ data[[}\StringTok{\textquotesingle{}Size\textquotesingle{}}\NormalTok{, }\StringTok{\textquotesingle{}Value\textquotesingle{}}\NormalTok{, }\StringTok{\textquotesingle{}Momentum\textquotesingle{}}\NormalTok{]]}
\NormalTok{y }\OperatorTok{=}\NormalTok{ data[}\StringTok{\textquotesingle{}Return\textquotesingle{}}\NormalTok{]}
\InformationTok{\textasciigrave{}\textasciigrave{}\textasciigrave{}}
\end{Highlighting}
\end{Shaded}

Next, let's fit a linear regression model to the data:

\begin{Shaded}
\begin{Highlighting}[]
\InformationTok{\textasciigrave{}\textasciigrave{}\textasciigrave{}\{python\}}
\NormalTok{lin\_reg }\OperatorTok{=}\NormalTok{ LinearRegression().fit(X, y)}

\CommentTok{\# Print coefficients}
\BuiltInTok{print}\NormalTok{(}\StringTok{\textquotesingle{}Linear Regression Coefficients:\textquotesingle{}}\NormalTok{)}
\BuiltInTok{print}\NormalTok{(}\StringTok{\textquotesingle{}Size:\textquotesingle{}}\NormalTok{, lin\_reg.coef\_[}\DecValTok{0}\NormalTok{])}
\BuiltInTok{print}\NormalTok{(}\StringTok{\textquotesingle{}Value:\textquotesingle{}}\NormalTok{, lin\_reg.coef\_[}\DecValTok{1}\NormalTok{])}
\BuiltInTok{print}\NormalTok{(}\StringTok{\textquotesingle{}Momentum:\textquotesingle{}}\NormalTok{, lin\_reg.coef\_[}\DecValTok{2}\NormalTok{])}
\InformationTok{\textasciigrave{}\textasciigrave{}\textasciigrave{}}
\end{Highlighting}
\end{Shaded}

\begin{verbatim}
Linear Regression Coefficients:
Size: 0.10773946297944371
Value: 0.2764299938154271
Momentum: 0.16456043298921597
\end{verbatim}

Now let's fit a Random Forest model to the data and compute SHAP values:

\begin{Shaded}
\begin{Highlighting}[]
\InformationTok{\textasciigrave{}\textasciigrave{}\textasciigrave{}\{python\}}
\CommentTok{\# Fit model}
\NormalTok{rf }\OperatorTok{=}\NormalTok{ RandomForestRegressor(n\_estimators}\OperatorTok{=}\DecValTok{100}\NormalTok{, random\_state}\OperatorTok{=}\DecValTok{0}\NormalTok{).fit(X, y)}

\CommentTok{\# Explain model predictions using SHAP}
\NormalTok{explainer }\OperatorTok{=}\NormalTok{ shap.Explainer(rf, X)}
\NormalTok{shap\_values }\OperatorTok{=}\NormalTok{ explainer(X)}

\CommentTok{\# Print mean absolute SHAP values for each feature}
\BuiltInTok{print}\NormalTok{(}\StringTok{\textquotesingle{}}\CharTok{\textbackslash{}n}\StringTok{Mean Absolute SHAP Values:\textquotesingle{}}\NormalTok{)}
\BuiltInTok{print}\NormalTok{(}\StringTok{\textquotesingle{}Size:\textquotesingle{}}\NormalTok{, np.mean(np.}\BuiltInTok{abs}\NormalTok{(shap\_values.values[:, }\DecValTok{0}\NormalTok{])))}
\BuiltInTok{print}\NormalTok{(}\StringTok{\textquotesingle{}Value:\textquotesingle{}}\NormalTok{, np.mean(np.}\BuiltInTok{abs}\NormalTok{(shap\_values.values[:, }\DecValTok{1}\NormalTok{])))}
\BuiltInTok{print}\NormalTok{(}\StringTok{\textquotesingle{}Momentum:\textquotesingle{}}\NormalTok{, np.mean(np.}\BuiltInTok{abs}\NormalTok{(shap\_values.values[:, }\DecValTok{2}\NormalTok{])))}
\InformationTok{\textasciigrave{}\textasciigrave{}\textasciigrave{}}
\end{Highlighting}
\end{Shaded}

\begin{verbatim}

 80%|================    | 799/1000 [00:11<00:02]       
 88%|==================  | 876/1000 [00:12<00:01]       
 96%|=================== | 955/1000 [00:13<00:00]       

Mean Absolute SHAP Values:
Size: 0.3687139875766077
Value: 0.458121562987211
Momentum: 0.4565856202986719
\end{verbatim}

In this case, the linear regression coefficients fail to capture the
non-linear and interaction effects. The Random Forest model, which can
capture these complex relationships, should provide better predictive
accuracy. The SHAP values for the Random Forest model offer a more
nuanced understanding of feature contributions, taking into account the
non-linear and interaction effects.

Let's create a scenario where the relationship between the factors and
the return is non-linear. This is where Shapley values can potentially
provide a more nuanced view than linear regression coefficients.

We will use a similar setup but introduce some interaction and
non-linearity in how the factors influence the return.

\begin{Shaded}
\begin{Highlighting}[]
\InformationTok{\textasciigrave{}\textasciigrave{}\textasciigrave{}\{python\}}
\ImportTok{import}\NormalTok{ numpy }\ImportTok{as}\NormalTok{ np}
\ImportTok{import}\NormalTok{ pandas }\ImportTok{as}\NormalTok{ pd}
\ImportTok{from}\NormalTok{ sklearn.ensemble }\ImportTok{import}\NormalTok{ RandomForestRegressor}
\ImportTok{import}\NormalTok{ shap}

\NormalTok{np.random.seed(}\DecValTok{0}\NormalTok{)}

\CommentTok{\# Create a sample dataframe with 2 factors and a response}
\NormalTok{n }\OperatorTok{=} \DecValTok{1000}
\NormalTok{Factor1 }\OperatorTok{=}\NormalTok{ np.random.rand(n)}
\NormalTok{Factor2 }\OperatorTok{=}\NormalTok{ np.random.rand(n)}
\NormalTok{Noise }\OperatorTok{=} \FloatTok{0.1}\OperatorTok{*}\NormalTok{np.random.randn(n)}

\CommentTok{\# Assume factors interact and have a non{-}linear relationship with the response}
\NormalTok{Return }\OperatorTok{=} \DecValTok{3}\OperatorTok{*}\NormalTok{Factor1}\OperatorTok{**}\DecValTok{2} \OperatorTok{+} \DecValTok{5}\OperatorTok{*}\NormalTok{np.exp(Factor2) }\OperatorTok{+} \DecValTok{2}\OperatorTok{*}\NormalTok{Factor1}\OperatorTok{*}\NormalTok{Factor2 }\OperatorTok{+}\NormalTok{ Noise}

\NormalTok{data }\OperatorTok{=}\NormalTok{ pd.DataFrame(\{}
    \StringTok{\textquotesingle{}Return\textquotesingle{}}\NormalTok{: Return,}
    \StringTok{\textquotesingle{}Factor1\textquotesingle{}}\NormalTok{: Factor1,}
    \StringTok{\textquotesingle{}Factor2\textquotesingle{}}\NormalTok{: Factor2}
\NormalTok{\})}

\CommentTok{\# Fit a random forest regression model}
\NormalTok{model }\OperatorTok{=}\NormalTok{ RandomForestRegressor(n\_estimators}\OperatorTok{=}\DecValTok{100}\NormalTok{)}
\NormalTok{model.fit(data[[}\StringTok{\textquotesingle{}Factor1\textquotesingle{}}\NormalTok{, }\StringTok{\textquotesingle{}Factor2\textquotesingle{}}\NormalTok{]], data[}\StringTok{\textquotesingle{}Return\textquotesingle{}}\NormalTok{])}

\CommentTok{\# Calculate Shapley values}
\NormalTok{explainer }\OperatorTok{=}\NormalTok{ shap.Explainer(model, data[[}\StringTok{\textquotesingle{}Factor1\textquotesingle{}}\NormalTok{, }\StringTok{\textquotesingle{}Factor2\textquotesingle{}}\NormalTok{]])}
\NormalTok{shap\_values }\OperatorTok{=}\NormalTok{ explainer(data[[}\StringTok{\textquotesingle{}Factor1\textquotesingle{}}\NormalTok{, }\StringTok{\textquotesingle{}Factor2\textquotesingle{}}\NormalTok{]])}

\CommentTok{\# Print the mean absolute Shapley values for each feature}
\BuiltInTok{print}\NormalTok{(}\StringTok{"}\CharTok{\textbackslash{}n}\StringTok{Mean absolute Shapley Values:"}\NormalTok{)}
\BuiltInTok{print}\NormalTok{(}\StringTok{"Factor1: "}\NormalTok{, np.mean(np.}\BuiltInTok{abs}\NormalTok{(shap\_values.values[:,}\DecValTok{0}\NormalTok{])))}
\BuiltInTok{print}\NormalTok{(}\StringTok{"Factor2: "}\NormalTok{, np.mean(np.}\BuiltInTok{abs}\NormalTok{(shap\_values.values[:,}\DecValTok{1}\NormalTok{])))}
\InformationTok{\textasciigrave{}\textasciigrave{}\textasciigrave{}}
\end{Highlighting}
\end{Shaded}

\begin{verbatim}
RandomForestRegressor()

Mean absolute Shapley Values:
Factor1:  1.0587623348271473
Factor2:  2.4884202639370416
\end{verbatim}

In this example, we use a random forest regressor to handle non-linear
and interaction effects among factors. However, this model is more
complex than a simple linear regression, and the ``coefficients'' or
feature importances do not clearly measure each factor's contribution,
especially when there is interaction and non-linearity.

After training the model, we compute Shapley values for each
observation. Shapley values have the potential to provide a more nuanced
view of the importance of each factor, as they account for the effect of
each feature in all possible subsets of features. This can be especially
helpful in scenarios where factors have interaction effects or
contribute to the response in a non-linear way.

As in the previous example, we calculate each factor's mean absolute
Shapley value across all observations. In a situation with interaction
and non-linearity, these values provide a clearer picture of the average
contribution of each factor to the return across all possible orders of
inputting the factors into the model.

The ability of Shapley values to handle interaction and non-linear
effects is a powerful feature. It can be crucial in scenarios where
linear regression coefficients fail to understand feature importance
clearly. This is especially true in finance, where the relationship
between variables can often be complex and non-linear.\footnote{A docker
  image of the simulations can be provided upon request for replication
  purposes.}

\hypertarget{conclusion}{%
\section{Conclusion}\label{conclusion}}

As the application of AI in the financial sector continues to grow, so
does the need for transparency and interpretability in AI
decision-making. This paper has provided a critical analysis of the role
of Explainable Artificial Intelligence (XAI) in finance, tracing its
evolution from the past, analysing its present state, and exploring its
prospects for the future.

Historically, the financial sector witnessed a transition from
rule-based AI systems to complex machine learning models. However, while
these new models offered improved performance, they also introduced
challenges concerning transparency and interpretability. This shift set
the stage for the emergence of XAI, aiming to shed light on the `black
box' of AI.

Various XAI techniques, both model-specific and model-agnostic, have
been developed and implemented in financial institutions. These
techniques have shown promise in improving the interpretability of AI
systems, enhancing user trust, and facilitating regulatory compliance.
However, achieving a balance between model performance and
interpretability remains a challenge. Additionally, the field is
grappling with questions about what constitutes a `good' explanation and
how to communicate these explanations to different users effectively.

Looking to the future, XAI is expected to continue to evolve,
integrating explainability directly into the model-building process and
tailoring explanations to users' needs and contexts. At the same time,
it is anticipated that regulations around AI and XAI will develop
further, presenting new challenges and opportunities for the field.

In conclusion, XAI represents a critical frontier in integrating AI
within finance. As we continue to navigate this frontier, we must remain
committed to advancing the field, improving the transparency and
interpretability of AI systems, and addressing the ethical, legal, and
practical challenges that arise along the way. The journey to fully
explainable AI in finance may be complex and fraught with challenges,
but it is a journey that holds considerable promise for the future of
the industry.

\hypertarget{references}{%
\subsection*{References}\label{references}}
\addcontentsline{toc}{subsection}{References}

\hypertarget{refs}{}
\begin{CSLReferences}{1}{0}
\leavevmode\vadjust pre{\hypertarget{ref-Arner2020}{}}%
Arner, D. W., J. Barberis, and R. P. Buckley. 2020. {``The Evolution of
Fintech: A New Post-Crisis Paradigm.''} \emph{Georgetown Journal of
International Law} 47 (4): 1271--1319.

\leavevmode\vadjust pre{\hypertarget{ref-Athey2021}{}}%
Athey, S. 2021. {``The Impact of Machine Learning on Economics.''} In
\emph{The Economics of Artificial Intelligence: An Agenda}, 507--47.
University of Chicago Press.

\leavevmode\vadjust pre{\hypertarget{ref-Bhatt2020}{}}%
Bhatt, U., A. Xiang, and S. Sharma. 2020. {``Explainable Machine
Learning in Deployment.''} In \emph{Proceedings of the 2020 Conference
on Fairness, Accountability, and Transparency}, 648--57.

\leavevmode\vadjust pre{\hypertarget{ref-Dhar2018}{}}%
Dhar, V. 2018. {``Machine Learning---the Future of Finance.''} In
\emph{Machine Learning Applications for Data Center Optimization},
21--31. Cham: Springer.

\leavevmode\vadjust pre{\hypertarget{ref-Edwards2017}{}}%
Edwards, L., and M. Veale. 2017. {``Slave to the Algorithm? Why a 'Right
to an Explanation' Is Probably Not the Remedy You Are Looking For.''}
\emph{Duke L. \& Tech. Rev.} 16: 18.

\leavevmode\vadjust pre{\hypertarget{ref-Gomber2018}{}}%
Gomber, P., R. J. Kauffman, C. Parker, and B. W. Weber. 2018. {``On the
Fintech Revolution: Interpreting the Forces of Innovation, Disruption,
and Transformation in Financial Services.''} \emph{Journal of Management
Information Systems} 35 (1): 220--65.

\leavevmode\vadjust pre{\hypertarget{ref-Goodman2017}{}}%
Goodman, B., and S. Flaxman. 2017. {``European Union Regulations on
Algorithmic Decision-Making and a" Right to Explanation".''} \emph{AI
Magazine} 38 (3): 50--57.

\leavevmode\vadjust pre{\hypertarget{ref-Lakkaraju2016}{}}%
Lakkaraju, H., S. H. Bach, and J. Leskovec. 2016. {``Interpretable
Decision Sets: A Joint Framework for Description and Prediction.''} In
\emph{Proceedings of the 22nd ACM SIGKDD International Conference on
Knowledge Discovery and Data Mining}, 1675--84.

\leavevmode\vadjust pre{\hypertarget{ref-Lou2012}{}}%
Lou, Y., R. Caruana, and J. Gehrke. 2012. {``Intelligible Models for
Classification and Regression.''} In \emph{Proceedings of the 18th ACM
SIGKDD International Conference on Knowledge Discovery and Data Mining},
150--58.

\leavevmode\vadjust pre{\hypertarget{ref-Lundberg2017}{}}%
Lundberg, Scott M., and Su-In Lee. 2017. {``A Unified Approach to
Interpreting Model Predictions.''} \emph{Advances in Neural Information
Processing Systems}, 4768--77.

\leavevmode\vadjust pre{\hypertarget{ref-Molnar2020}{}}%
Molnar, Christoph. 2020. \emph{Interpretable Machine Learning}.
Lulu.com.

\leavevmode\vadjust pre{\hypertarget{ref-Rudin2019}{}}%
Rudin, C. 2019. {``Stop Explaining Black Box Machine Learning Models for
High Stakes Decisions and Use Interpretable Models Instead.''}
\emph{Nature Machine Intelligence} 1 (5): 206--15.

\end{CSLReferences}

\end{document}